\documentclass[journal]{IEEEtran}
\usepackage{cite}
\usepackage{amsmath,amssymb,amsfonts}
\usepackage{bm,bbm}
\usepackage[noend]{algorithmic}
\usepackage{graphicx,color}
\usepackage{textcomp}
\usepackage{xcolor}
\usepackage{hyperref}
\usepackage{comment}
\usepackage{siunitx}
\usepackage{placeins}
\usepackage{multirow}
\hypersetup{hidelinks=true}
\usepackage{algorithm}
\usepackage[nolist]{acronym}

\begin{acronym}
\acro{6G}{sixth generation}
\acro{5G}{fifth generation}
\acro{AP}{access point}
\acro{AWGN}{additive white Gaussian noise}
\acro{CDF}{cumulative distribution function}
\acro{CF-MaMIMO}{cell-free massive multiple-input multiple-output}
\acro{CPU}{central processing unit}
\acro{CSI}{channel state information}
\acro{EP}{expectation propagation}
\acro{EVD}{eigenvalue decomposition}
\acro{GTA}{Gaussian tree approximation}
\acro{ICD}{iterative channel estimation and data detection}
\acro{JCD}{joint channel estimation and data detection}
\acro{KL}{Kullback-Leibler}
\acro{MAC}{multiple access channel}
\acro{MaMIMO}{massive multiple-input multiple-output}
\acro{MAP}{maximum-a-posteriori}
\acro{MIMO}{multiple-input multiple-output}
\acro{MMSE}{minimum mean squared error}
\acro{NMSE}{normalized mean squared error}
\acro{PDF}{probability density function}
\acro{PMF}{probability mass function}
\acro{QoS}{quality of  service}
\acro{SER}{symbol error rate}
\acro{SIC}{successive interference cancellation}
\acro{SIMO}{single-input multiple-output}
\acro{SINR}{signal-to-interference-and-noise-ratio}
\acro{SISO}{soft-input soft-output}
\acro{UE}{user equipment}
\acro{UL}{uplink}
\acro{ZF}{zero forcing}
\end{acronym}

\newcommand{\lvec}[1]{\ensuremath{\bm{#1}}}  		
\newcommand*{\argmin}{\ensuremath{\mathop{\mathrm{arg\,min}}}}

\newcommand{\est}[2]{\ensuremath{E_{#1}\!\left\{#2\right\}}}
\newcommand{\estemp}[1]{\ensuremath{\hat{E}\!\left\{#1\right\}}}
\newcommand{\msgn}{\ensuremath{m}}

\newcommand{\msg}[2]{\ensuremath{m_{#1;#2}}}

\newcommand{\qmsg}[2]{\ensuremath{q_{#1;#2}}}

\newcommand{\mumsg}[2]{\ensuremath{\bm{\mu}_{#1;#2}}}
\newcommand{\cmsg}[2]{\ensuremath{\bm{C}_{#1;#2}}}

\begin{document}

\title{Bilinear Expectation Propagation for Distributed Semi-Blind Joint Channel Estimation and Data Detection in Cell-Free Massive MIMO}

\author{Alexander~Karataev,
        Christian~Forsch,~\IEEEmembership{Graduate Student Member,~IEEE,} and~Laura~Cottatellucci,~\IEEEmembership{Member,~IEEE}
        \thanks{This work was funded by the Deutsche Forschungsgemeinschaft (DFG, German Research Foundation) – Project CO 1311/1-1,  Project ID 491320625.}
        \thanks{Alexander Karataev, Chrstian Forsch, and Laura Cottatellucci are with the Institute for Digital Communications, Friedrich-Alexander-Universität Erlangen-Nürnberg, Erlangen, Germany (e-mail: alexander.karataev@fau.de; christian.forsch@fau.de; laura.cottatellucci@fau.de).}}
        
\maketitle

\begin{abstract}
We consider a \ac{CF-MaMIMO}  communication system in the uplink transmission and propose a novel algorithm for blind or semi-blind \ac{JCD}. We formulate the problem in the framework of bilinear inference and develop a solution based on the \ac{EP} method for both channel estimation and data detection.  We propose a new approximation of the joint a posteriori distribution of the channel and data whose representation as a factor graph enables the application of the \ac{EP} approach using the message-passing technique, local low-complexity computations at the nodes, and an effective modeling of channel-data interplay.  The derived algorithm, called bilinear-EP \ac{JCD},  allows for a distributed implementation among \acp{AP} and the \ac{CPU} and has polynomial complexity. Our simulation results show that it outperforms other \ac{EP}-based state-of-the-art polynomial time algorithms.
\end{abstract}



\maketitle

\section{Introduction}\label{intro}

\IEEEPARstart{C}{ell-free} massive multiple-input multiple-output (CF-MaMIMO)\acused{CF-MaMIMO}\acused{MIMO}\acused{MaMIMO} networks enable primary goals for \ac{6G} wireless communication systems, such as ubiquitous coverage with uniform  \ac{QoS} and ultra-high-rate, energy-efficient data transmission~\cite{Ngo2017,Ngo2018,Yang2018, Ammar2022}. In \ac{CF-MaMIMO} systems, a large number of geographically distributed \acp{AP} are jointly serving a much smaller number of \acp{UE}.  The joint processing is coordinated by one or multiple \acp{CPU} which are connected to the \acp{AP} via fronthaul links.
The geographically distributed nature of \ac{CF-MaMIMO} networks enhances the attractive properties of centralized \ac{MaMIMO} systems by reducing the average minimum  distance between \acp{AP} and \acp{UE}. This allows \ac{CF-MaMIMO} networks to provide uniform high data rates over the coverage area and high energy efficiency. However, in contrast to centralized \ac{MaMIMO},  channel hardening and favorable propagation typically do not hold \cite{Yin2014,Chen2018,Gholami2020a,Gholami2020b}. Thus, the low-complexity matched filter, which provides near-optimal detection performance in \ac{MaMIMO}  systems \cite{Marzetta2010},
is not effective  for \ac{CF-MaMIMO} systems, and optimal joint signal processing at the centralized \ac{CPU} is not computationally affordable. Similarly,  existing pilot decontamination solutions for centralized \ac{MaMIMO}~\cite{Ngo2012,Yin2013,Cottatellucci2013,Mueller2014,Yin2016} are not effective.
Thus, the quest for low-complexity detection techniques with performance approaching that of centralized joint optimal processing schemes is still an open challenge as well as  the search of effective pilot decontamination methods.

Distributed processing at the \acp{AP} can efficiently reduce computational complexity at the \ac{CPU}, see e.g., the extensive analysis in \cite{Bjoernson2020, Demir2021} and \cite{Wang2020}, whereas semi-blind channel estimation has shown  to combat effectively pilot contamination \cite{Cottatellucci2013,Mueller2014,Yin2016}. Therefore, in this work, we propose a distributed semi-blind \ac{JCD} algorithm based on \ac{EP} which exhibits a low complexity.

\ac{EP} is an approximate inference technique which iteratively finds a tractable approximation of factorized probability distributions by projecting each factor onto an exponential family~\cite{Minka2001a,Minka2001b}.
\ac{EP} was already applied  for data detection in previous works.
In~\cite{Cespedes2014}, the authors proposed a low-complexity \ac{EP}-based \ac{MIMO} detector while assuming perfect \ac{CSI} and  a Gaussian approximation for the posterior distribution of data symbols. 
In~\cite{Ghavami2017b}, the authors extended the EP-detection algorithm proposed in~\cite{Cespedes2014} to imperfect \ac{CSI} by embedding channel estimation errors in the \ac{EP} formulation. 

The \ac{EP} method was also applied to blind channel estimation and noncoherent detection. In~\cite{Ghavami2018}, the authors presented a blind channel estimation algorithm for multi-cell \ac{MaMIMO} systems. In~\cite{Ngo2020},  a noncoherent multi-user detection scheme was proposed for the \ac{SIMO} \ac{MAC}.  In both schemes, the approximated channel and symbol distributions were chosen to be multivariate Gaussian  and multinoulli  distributions, respectively. In~\cite{Ghavami2018},  the proposed approximate joint posterior distribution of channels and transmitted symbols resulted in two \emph{decoupled} \ac{EP}-based schemes for channel estimation and data detection but, because of the high complexity of the \ac{EP}-based symbol detection, only  \ac{EP}-based channel estimation was retained for practical system implementations followed by  conventional linear symbol detectors. 
The decoupling between channel estimation and symbol detection implied that the   \ac{EP}-based channel estimation could only exploit  prior statistical knowledge on the transmitted symbols but not their deterministic imperfect knowledge.   
In~\cite{Ngo2020}, the proposed approximate posterior distribution factorization yielded a factor tree with a branch per user.  The detection scheme was derived by applying message-passing rules for \ac{EP} on this tree. The resulting algorithm could be applied to both pilot-assisted and pilot-free communications. However, the complexity of the proposed algorithm was exponential in the product of the number of channel uses and the logarithm of the symbol constellation set cardinality and, hence, it was not affordable for practical high data rate systems.

The \ac{EP} framework was also used to develop decentralized detection schemes for \ac{MaMIMO} systems in \cite{Wang2020, Zhang2020, Dong2022, Li2023}. In these works,  the computational complexity  was reduced compared to centralized schemes by processing  the signals received by  antenna sub-arrays locally  via \ac{EP} message passing and then combining the messages from sub-arrays at the \ac{CPU}. The posterior data symbol distributions were approximated by multivariate Gaussians and  perfect \ac{CSI} knowledge was assumed. In \cite{Wang2020}, the authors  reduced  message sizes by utilizing averaging.  In~\cite{Zhang2020}, the decentralized \ac{MaMIMO} receiver embedded both detection and decoding and the  extrinsic information from the decoder was utilized as a priori information for the \ac{EP}-based detector.
In~\cite{Dong2022},  the authors introduced a pre-processing based on QR-factorization and a variance compensation scheme in the decentralized detector of~\cite{Zhang2020}.
In~\cite{Li2023}, two decentralized \ac{EP} detection approaches were proposed, the first based on user grouping and, thus, group-wise joint detection, and the second on a daisy-chain architecture. 

\ac{EP}-based receivers were also applied in \ac{CF-MaMIMO} systems.
The authors of~\cite{Kosasih2021} utilized a centralized \ac{EP}-based detector with Gaussian data approximations which incorporates channel estimation errors as in~\cite{Ghavami2017b}.
The channel estimation error accounted for pilot contamination and general estimation errors due to noise.
In~\cite{He2021}, the authors proposed a distributed \ac{EP} detector for \ac{CF-MaMIMO} based on the decentralized subarray-based detector in~\cite{Wang2020}.
The aforementioned approach was extended to an \ac{ICD} scheme in~\cite{He2023}.
Here, the data detection was based on \ac{EP} and the iterative algorithm took into account the channel estimation errors. The channel estimation was based on \ac{MMSE} estimation with detected data symbols as additional pilots.

In this work, we propose a new distributed algorithm for \ac{JCD} in \ac{CF-MaMIMO} systems.
Our contributions can be summarized as follows:
\begin{itemize}
	\item We develop a novel message-passing algorithm for a bilinear inference problem arising in \ac{JCD}.  The inference is based on the approximate \ac{EP} method and assumes general multivariate Gaussian and multinoulli distributions for the posterior distributions  of the \ac{AP} channels and  data symbols, respectively, as in \cite{Ngo2020}.  In contrast to the approximate posterior distributions of \cite{Ghavami2018, Ngo2020}, the factorization of the proposed approximate posterior joint distribution for channels and data symbols allows for an alternating refinement of channel and data estimates and  a distributed implementation of the algorithm with local processing at the \acp{AP}. Furthermore, the inclusion of single-user \ac{SISO} decoders is straightforward. Finally, the algorithm can automatically take into account the effect of pilot contamination\footnote{The analysis of the impact of \ac{SISO} decoding and pilot contamination exceeds the  scope of this paper.}.
 \item We show that the proposed bilinear-EP \ac{JCD} algorithm has polynomial complexity in system parameters and bridges the complexity gap between algorithms based on \ac{EP} that approximate the posterior distribution of data with a Gaussian distribution such as, e.g., \cite{ Cespedes2014, Ghavami2017a, Ghavami2018, Wang2020, Zhang2020, Dong2022, Li2023, Kosasih2021, He2021, He2023} and those that assume more precise categorical distributions~\cite{Ngo2020}.
 	\item We consider four baseline schemes, namely a centralized linear \ac{MMSE} detector, the detector in \cite{He2021} assuming perfect \ac{CSI}, the \ac{ICD} algorithm in \cite{He2023}, and a modified version of the proposed \ac{JCD} algorithm with perfect channel knowledge which provides a lower bound to the \ac{SER}. Our simulation results show the superior performance of our approach compared to the first three baseline algorithms.
\end{itemize}

It is worth noting that the appealing features of the proposed algorithm stem from the choice of the \emph{factorization} of the approximate joint posterior distribution and the way to take  into account the interplay between the two sets of variables, i.e., channel coefficients and  data symbols. This method can be applied  to other bilinear inference problems such as gain calibration, matrix factorization and compressed matrix sensing~\cite{schulke2016statistical}. 
   
This paper is organized as follows.
In Section~\ref{sys_mod}, we introduce the \ac{CF-MaMIMO} system model.
In Section~\ref{EP_Graphs}, we review the \ac{EP} algorithm and its application on graphs.
Based on this theoretical framework, we develop the bilinear-EP \ac{JCD} algorithm in Section~\ref{delta-EP}.
Then, in Section~\ref{sim_res}, we present numerical results which show the superior performance of the proposed algorithm.
Finally, some conclusions are drawn.

\textit{Notation:}
Lower case, bold lower case, and bold upper case letters,  e.g.,  $x, \bm{x}, \bm{X},$ represent scalars, vectors, and matrices, respectively.
$\bm{I}_N$ is the identity matrix of dimension $N$.
$\{x_{l,k,t}\}=\bigcup_{l,k,t}x_{l,k,t}$ denotes the collection of all variables obtained by varying the possible indices. $\mathcal{A}\setminus \mathcal{B}$ is the set complement operator between two sets $\mathcal{A},\mathcal{B}$. $\delta(\cdot)$ is the Dirac delta function. The indicator function $\mathbbm{1}_{x\in\mathcal{S}}$ takes value $1$ if the condition in the subscript is satisfied and zero otherwise, e.g. element $x$ is in the set $\mathcal{S}$. The operator  $\mathrm{vec}(\bm{X})$ maps the matrix $\bm{X}$ onto the vector  obtained by stacking the columns of $\bm{X}$ on top of one another. $\otimes$ is the Kronecker product between two matrices. $\est{p(\bm{x})}{\cdot}$ stands for the expectation operator with respect to the probability distribution $p(\bm{x})$ and $\estemp{\cdot}$ stands for the empirical expectation operator. We denote by $\pi(\cdot)$ and  $\mathcal{N}(\bm{x};\bm{\mu},\bm{C})$  the categorical distribution and the circularly symmetric Gaussian distribution of complex-valued vectors $\bm{x}$ with mean vector $\bm{\mu}$ and covariance matrix $\bm{C}$, respectively.
The notation $x \sim p$ indicates  that the random variable $x$ follows the distribution $p$.

\section{System Model}\label{sys_mod}
We consider a communication system in the uplink transmission that consists of $L$ \acp{AP} and $K$ single-antenna \acp{UE}. Each \ac{AP} is equipped with $N$ co-located antennas. Each user sends a signal vector $\overline{\bm{x}}_k=[\bm{x}_{p;k}^T,\bm{x}_{k}^T]^T$ consisting of a $P$-dimensional pilot vector $\bm{x}_{p;k}\in\mathbb{C}^{P}$ and a $T$-dimensional data vector $\bm{x}_{k}\in\mathcal{S}^{T}$ where $\mathcal{S}$ is the employed constellation set. We combine pilot and data vectors of all users in the signal matrix $\overline{\bm{X}}=[\overline{\bm{x}}_1,\overline{\bm{x}}_2,...,\overline{\bm{x}}_K]^T\in\mathbb{C}^{K\times(P+T)}$. The matrix $\overline{\bm{X}}$ can also  be expressed as $\overline{\bm{X}}=[\bm{X}_p,\bm{X}]$ where $\bm{X}_p$ and $\bm{X}$ are matrices consisting of the pilot and data vectors of all users, respectively.
The equivalent complex baseband received signal at \ac{AP} $l\in\{1,...,L\}$ is given by
\begin{equation}
    \label{System Model Equation}
	\overline{\bm{Y}}_l=[\bm{Y}_{p;l},\bm{Y}_l]=\bm{H}_l[\bm{X}_p,\bm{X}]+\bm{N}_l,
\end{equation}
where $\overline{\bm{Y}}_l\in\mathbb{C}^{N\times(P+T)}$ denotes a matrix whose element $\overline{y}_{l,i,j}$ is the received signal at antenna $i$ of \ac{AP} $l$ during the $j$-th channel use.  Here, $\bm{Y}_{p;l}$ and $\bm{Y}_l$ are the components of the matrix $\overline{\bm{Y}}_l$ corresponding to the pilot matrix $\bm{X}_p$ and the data matrix $\bm{X}$, respectively. $\bm{H}_l\in\mathbb{C}^{N\times K}$ is the block Rayleigh fading channel matrix whose element $h_{l, n,k}$ is the coefficient of the  channel between user $k$ and antenna $n$ at \ac{AP} $l$, which is assumed to be constant during $P+T$ consecutive channel uses, and $\bm{N}_l\in\mathbb{C}^{N\times(P+T)}$ is a  matrix whose element $n_{l,i,j}$ is the \ac{AWGN} at antenna $i$ during the $j$-th channel use at the $l$-th \ac{AP}.
Therefore, both channel and additive noise are zero mean complex Gaussian vectors with covariance matrices $\bm{\Xi}_l\in\mathbb{C}^{NK\times NK}$ and $\sigma^2 \bm{I}_{N(P+T)}$, respectively, i.e., $\mathrm{vec}(\bm{H}_l)\sim\mathcal{N}(\bm{0},\bm{\Xi}_l)$ and $\mathrm{vec}(\bm{N}_l)\sim\mathcal{N}(\bm{0},\sigma^2 \bm{I}_{N(P+T)}).$
Furthermore, we assume that  the channels of different users are uncorrelated. Thus, the covariance matrix $\bm{\Xi}_l$ is a block diagonal matrix of $K$ blocks  $\bm{\Xi}_{kl}\in\mathbb{C}^{N\times N}$.

\section{Expectation Propagation on Graphs}\label{EP_Graphs}
In the following, we associate a factor graph to the factorized true distribution and describe a message-passing scheme that results from the local computation of the approximate factors on the factor graph nodes. 
More details on the derivation of the message-passing update rules can be found in~\cite{Minka2005,Heskes2005,Ngo2020}.

We aim at computing an approximation of a joint probability distribution which is assumed to be factorizable as follows
\begin{equation}
p(\lvec{x})=\prod_\alpha \Psi_\alpha(\lvec{x}_\alpha),
	\label{eq:posterior_factorization}
\end{equation}
where  $\lvec{x}_\alpha$ is a sub-vector of $\lvec{x},$ i.e., $\lvec{x}_\alpha\subseteq\lvec{x},$ and $\bigcup_\alpha \lvec{x}_\alpha = \lvec{x}$.
The approximation $\hat{p}(\lvec{x})$ of the joint distribution reflects the same factorized form,
\begin{equation}
	\hat{p}(\lvec{x}) = \frac{1}{Z}\cdot\prod_\alpha \hat{p}_\alpha(\lvec{x}_\alpha),
	\label{eq:approx_factorization}
\end{equation}
where $Z$ is a normalization constant and  $\hat{p}$ is constrained to lie within the exponential family $\mathcal{F}$.
Furthermore, we assume that the  approximation can be  fully factorized as follows
\begin{equation}
	\hat{p}(\lvec{x}) = \prod_\beta \hat{p}_\beta(\lvec{x}_\beta),
	\label{eq:approx_fully_factorized}
\end{equation}
where the sub-vectors $\lvec{x}_\beta$ contain elements  which always occur together in a factor.
The sub-vectors $\lvec{x}_\beta$ are not overlapping, i.e., $\lvec{x}_\beta \cap \lvec{x}_{\beta'} = \emptyset\quad\forall\beta\neq\beta'$. Furthermore, we have $\bigcup_{\beta}\bm{x}_{\beta}=\bm{x}$. Therefore, the sub-vectors $\bm{x}_{\beta}$ define a partition of $\bm{x}$.
Additionally, it holds for all sub-vectors $\bm{x}_{\beta}$ and $\bm{x}_{\alpha}$ that the sub-vector $\bm{x}_{\beta}$ either lies fully within $\bm{x}_{\alpha}$, i.e., $\bm{x}_{\beta}\subseteq\bm{x}_{\alpha}$, or they are completely disjoint, i.e., $\bm{x}_{\beta}\cap\bm{x}_{\alpha}=\emptyset$.

\begin{figure*}[tbp]
\normalsize
\centering
	\includegraphics[width=1\linewidth]{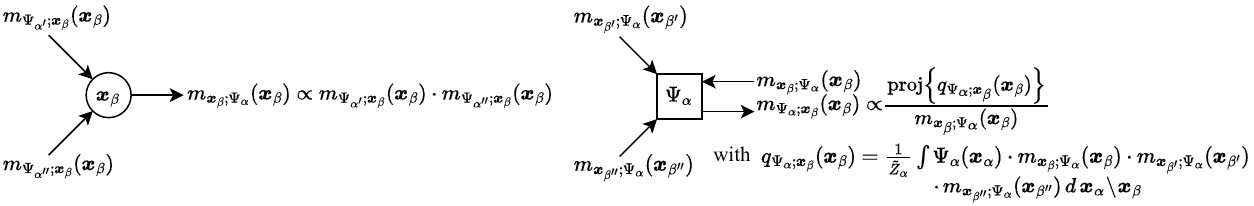}
	 \caption{Illustration of the \ac{EP} message-passing update rules.}
	\label{fig:EP_MP_updates}
\hrulefill
\end{figure*}
The factorization described above induces a representation of the joint distribution by a factor graph with factor nodes associated to functions  $\Psi_\alpha$ and variable nodes associated to sub-vectors $\lvec{x}_\beta$.  An edge exists between factor node  $\Psi_\alpha$ and variable node $\lvec{x}_\beta$ if $\bm{x}_{\beta}\subseteq\bm{x}_{\alpha}.$
We define the neighbor set $N_\alpha$ of factor node $\Psi_\alpha$ as the set of indices $\beta$ of all variable nodes $\lvec{x}_\beta$ that are connected to $\Psi_\alpha$, i.e., $N_\alpha=\{\beta\,|\,\lvec{x}_\beta\subseteq\lvec{x}_\alpha\}$.
Similarly, we define the neighbor set $N_\beta$ of variable node $\lvec{x}_\beta$ as the set of indices $\alpha$ of all factor nodes $\Psi_\alpha$ that are connected to $\lvec{x}_\beta$, i.e., $N_\beta=\{\alpha\,|\,\lvec{x}_\beta\subseteq\lvec{x}_\alpha\}$.

Since the approximation $\hat{p}$ lies within the exponential family $\mathcal{F}$, the approximate factors $\hat{p}_\alpha$ and $\hat{p}_\beta$ belong also to $\mathcal{F}$.
The assumption of the fully-factorized approximation in~\eqref{eq:approx_fully_factorized} yields the following factorizations for the approximate factors,
\begin{align}
	\hat{p}_\alpha(\lvec{x}_\alpha) &= \prod_{\beta\in N_\alpha} \msg{\Psi_\alpha}{\lvec{x}_\beta}(\lvec{x}_\beta),
	\label{eq:approx_factor_alpha_factorization}\\
	\hat{p}_\beta(\lvec{x}_\beta) &= \frac{1}{Z_\beta}\prod_{\alpha\in N_\beta} \msg{\Psi_\alpha}{\lvec{x}_\beta}(\lvec{x}_\beta),
	\label{eq:approx_factor_beta_factorization}
\end{align}
where $Z_\beta$ is a normalization constant and $\msg{\Psi_\alpha}{\lvec{x}_\beta}(\lvec{x}_\beta)$ is interpreted as a message from the factor node $\Psi_\alpha$ to the variable node $\lvec{x}_\beta$  also belonging to the specified exponential family.
Thus, we can update the approximate factors, and therefore also the approximate joint distribution $\hat{p}(\lvec{x})$, by exchanging and updating messages on a factor graph.

The \ac{EP} algorithm is an iterative algorithm.
Hence, we denote the messages during iteration $i$ as $\msgn^{(i)}$.
At first, we initialize all the factor-to-variable messages $\msg{\Psi_\alpha}{\lvec{x}_\beta}^{(0)}(\lvec{x}_\beta)$. For initialization, prior statistical  knowledge or, if not available,  uninformative distributions can be used. A particular example is discussed in section \ref{delta-EP}.
Then, the variable-to-factor messages and the factor-to-variable messages are updated, respectively, as follows,
\begin{align}
	\msg{\lvec{x}_\beta}{\Psi_\alpha}^{(i-1)}(\lvec{x}_\beta) &\propto \prod_{\alpha'\in N_\beta\setminus\alpha}\msg{\Psi_{\alpha'}}{\lvec{x}_\beta}^{(i-1)}(\lvec{x}_\beta),
	\label{eq:EP_var_to_fac_message}\\
	\msg{\Psi_\alpha}{\lvec{x}_\beta}^{(i)}(\lvec{x}_\beta) &\propto \frac{\text{proj}\left\{\qmsg{\Psi_{\alpha}}{\lvec{x}_{\beta}}^{(i-1)}(\lvec{x}_\beta)\right\}}{\msg{\lvec{x}_\beta}{\Psi_\alpha}^{(i-1)}(\lvec{x}_\beta)},
	\label{eq:EP_fac_to_var_message}
\end{align}
where the distribution $\qmsg{\Psi_{\alpha}}{\lvec{x}_{\beta}}^{(i-1)}(\lvec{x}_\beta)$ in~\eqref{eq:EP_fac_to_var_message} is given by
\begin{equation}
	\qmsg{\Psi_{\alpha}}{\lvec{x}_{\beta}}^{(i-1)}(\lvec{x}_\beta) = \frac{1}{\tilde{Z}_\alpha}\int\Psi_\alpha(\lvec{x}_\alpha)\prod_{\beta'\in N_\alpha}\msg{\lvec{x}_{\beta'}}{\Psi_\alpha}^{(i-1)}(\lvec{x}_{\beta'})\,\mathrm{d}\lvec{x}_\alpha\!\setminus\!\lvec{x}_\beta.
	\label{eq:message_projection}
\end{equation}
Here, $\tilde{Z}_\alpha$ is a normalization constant and $\text{proj}\{\cdot\}$ denotes the projection operator defined as
\begin{equation}
	\text{proj}\{f(\lvec{x})\} = \argmin_{g(\lvec{x})\in\mathcal{F}}D_{KL}(f(\lvec{x})||g(\lvec{x})),
	\label{eq:projection}
\end{equation}
where $D_{KL}(f(\lvec{x})||g(\lvec{x}))$ is the \ac{KL} divergence between $f$ and $g$ and $\mathcal{F}$ is an exponential family distribution.
The message-passing update rules~\eqref{eq:EP_var_to_fac_message} and~\eqref{eq:EP_fac_to_var_message} are illustrated in Fig.~\ref{fig:EP_MP_updates}.

\section{Bilinear-EP JCD}\label{delta-EP}
\subsection{Factorization and Messages}
\begin{figure}
		\includegraphics[width=\linewidth]{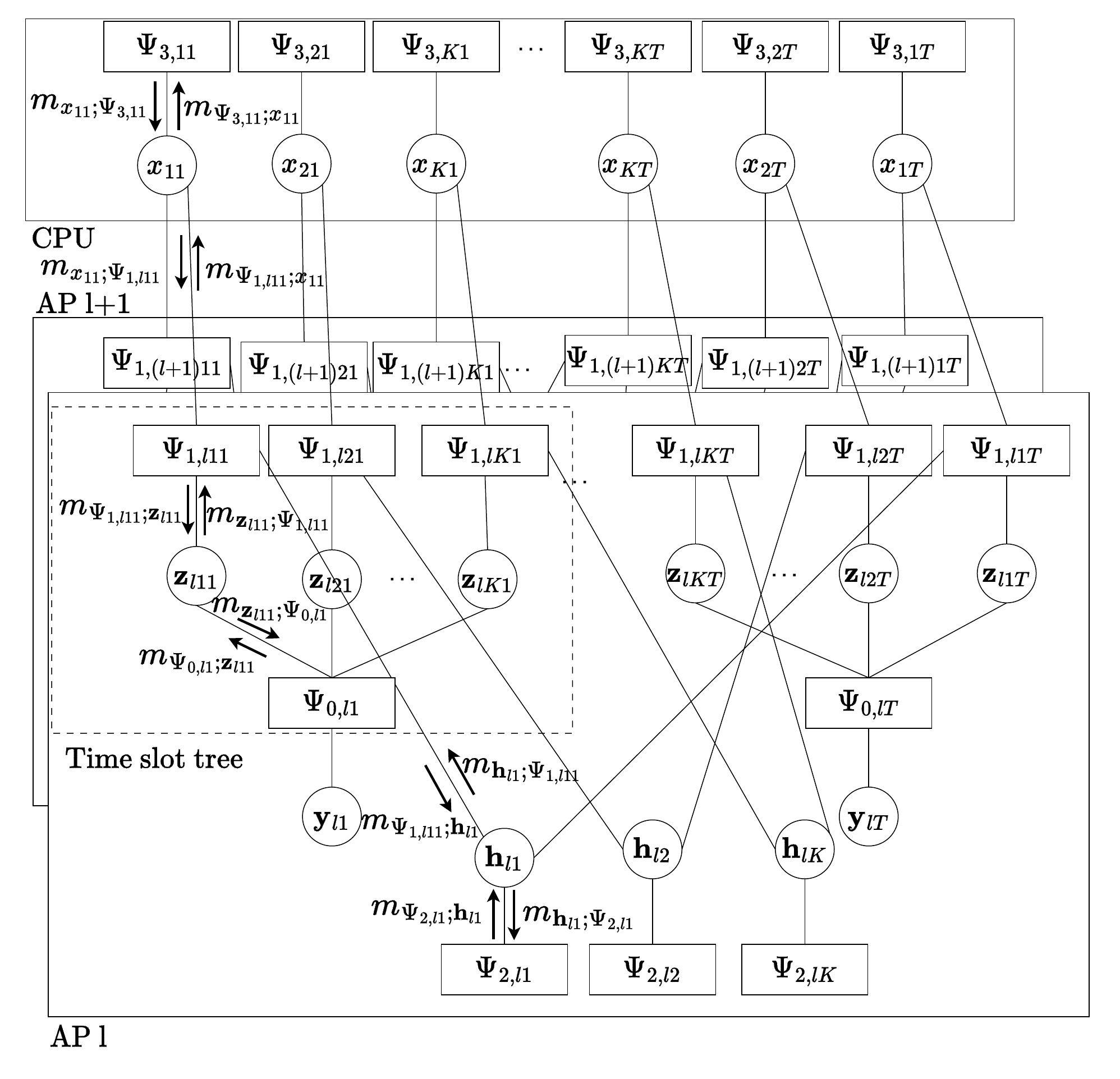}
		\caption{Factor graph of bilinear-EP \ac{JCD} for distributed \ac{CF-MaMIMO}.}
		\label{fig:Delta_JCD_L_AP}
\end{figure}

In the following, we derive an EP-based algorithm for semi-blind JCD which we call bilinear-EP \ac{JCD}. In this section, we focus on data signals since the received pilot signals are utilized to characterize the a priori channel distribution.
Inspired by \cite{Ngo2020}, we propose a novel factorization of the joint posterior distribution of data and channel. We decompose \eqref{System Model Equation} in $T$ vector relations, one for each channel use $t$, as follows
\begin{equation}
    \label{Output_Z_Relation}
    \bm{y}_{lt}=\sum_{k=1}^K x_{kt}\bm{h}_{lk}+\bm{n}_{lt},
\end{equation}
where $\bm{h}_{lk}$ is the $k$-th column, $\bm{y}_{lt}$ and $\bm{n}_{lt}$ are the $t$-th column and $x_{kt}$ is the element in row $k$ and column $t$ of the matrices $\bm{H}_l$, $\bm{Y}_l$, $\bm{N}_l$ and $\bm{X}$ respectively. 
Furthermore, we define
\begin{equation}
    \label{Z_Channel_Symbol_Relation}
    \bm{z}_{lkt}:=x_{kt}\bm{h}_{lk}.
\end{equation}
We can factorize the posterior density function using Bayes theorem as follows
\begin{align}\nonumber
	&p_{\{x_{kt}\},\{\bm{z}_{lkt}\},\{\bm{h}_{lk}\}|\{\bm{y}_{lt}\},\{\bm{Y}_{p;l}\}}\nonumber\\
 &\propto p_{\{x_{kt}\},\{\bm{z}_{lkt}\},\{\bm{h}_{lk}\},\{\bm{y}_{lt}\},\{\bm{Y}_{p;l}\}}\nonumber\\
 &=p_{\{\bm{y}_{lt}\}|\{\bm{z}_{lkt}\}}p_{\{\bm{z}_{lkt}\}|\{x_{kt}\},\{\bm{h}_{lk}\}}p_{\{x_{kt}\}}p_{\{\bm{h}_{lk}|\bm{Y}_{p;l}\}}\label{Posterior_Density_after_Bayes},
\end{align}
where the last equality follows from \eqref{Output_Z_Relation} and \eqref{Z_Channel_Symbol_Relation}.
We further assume that the a priori channel distributions are independent across users $k$ and \acp{AP} $l$. Then, exploiting  the independence of the noise across time $t$ and \acp{AP} $l$, the independence of the data symbols $\{x_{kt}\}$, and \eqref{Z_Channel_Symbol_Relation}, we can factorize \eqref{Posterior_Density_after_Bayes}  as follows
\begin{equation}
\begin{split}
&p_{\{x_{kt}\},\{\bm{z}_{lkt}\},\{\bm{h}_{lk}\}|\{\bm{y}_{lt}\},\{\bm{Y}_{p;l}\}}\\
&\propto\prod_{l=1}^{L}\left[\prod_{t=1}^{T}\left(p_{\bm{y}_{lt}|\bm{z}_{l1t},...,\bm{z}_{lKt}}\prod_{k=1}^{K}p_{\bm{z}_{lkt}|x_{kt},\bm{h}_{lk}}\right)\right]\\
&\times \prod_{k=1}^{K}\prod_{l=1}^{L}p_{\bm{h}_{lk}|\bm{Y}_{p;l}}\prod_{k=1}^{K}\prod_{t=1}^{T}p_{x_{kt}}\label{Posterior_Density_after_Bayes_fact},
\end{split}
\end{equation} 
For the probability distributions in~\eqref{Posterior_Density_after_Bayes_fact} we adopt the following compact notation
\begin{align*}
\Psi_{0,lt} &:=p_{\bm{y}_{lt}|\bm{z}_{l1t},...,\bm{z}_{lKt}},\\
\Psi_{1,lkt} &:=p_{\bm{z}_{lkt}|x_{kt},\bm{h}_{lk}},\\
\Psi_{2,lk} &:=p_{\bm{h}_{lk}|\bm{Y}_{p;l}},\\
\Psi_{3,kt} &:=p_{x_{kt}}.
\end{align*}
The factor graph induced by the factorization in (\ref{Posterior_Density_after_Bayes_fact}) is illustrated in Fig.~\ref{fig:Delta_JCD_L_AP} and contains cycles.
According to the system model in (\ref{System Model Equation}) the true factors are given by
\begin{align*}
\Psi_{0,lt} &= \mathcal{N}(\bm{y}_{lt};\sum_{k=1}^{K}\bm{z}_{lkt},\sigma^2\bm{I}_N),\\
\Psi_{1,lkt}&=\delta(\bm{z}_{lkt}-x_{kt}\bm{h}_{lk}),\\
\Psi_{2,lk} &=\mathcal{N}(\bm{h}_{lk},\bm{\mu}_{lk;\mathrm{MMSE}},\bm{C}_{lk;\mathrm{MMSE}}),\\
\Psi_{3,kt}&=\frac{1}{\vert\mathcal{S}\vert}\mathbbm{1}_{x_{kt}\in\mathcal{S}},
\end{align*}
where we assume uniform prior distributions  $p_{x_{kt}}$ for data symbols and $\Psi_{2,lk}$ is given by the pilot-based Bayesian \ac{MMSE} channel estimate of the channel $\bm{H}_l$ as follows
\begin{align*}
p_{\bm{H}_l|\bm{Y}_{p;l}}=\mathcal{N}\left(\mathrm{vec}(\bm{H}_l);\frac{1}{\sigma^2}\bm{C}\bm{A}_p^H\mathrm{vec}(\bm{Y}_{p;l}),\bm{C}\right)
\end{align*}
with   ${\bm{C}:=\left(\bm{\Xi}_l^{-1}+\frac{1}{\sigma^2}\bm{A}_p^H\bm{A}_p\right)^{-1}}$ and $\bm{A}_p:=(\bm{X}_p^T\otimes\bm{I}_N)$. 
Due to the independence of the channels among different users, we have
$\bm{\mu}_{lk;\mathrm{MMSE}}=\left[\frac{1}{\sigma^2}\bm{C}\bm{A}_p^H\mathrm{vec}(\bm{Y}_{p;l})\right]_{(k-1)N+1:kN}$  and  $    \bm{C}_{lk;\mathrm{MMSE}}=\bm{C}_{(k-1)N+1:kN,(k-1)N+1:kN}.$
We choose the approximate factor-to-variable messages to be distributed as
\begin{align*}
	\msg{\Psi_{0,lt}}{\bm{z}_{lkt}}&=\mathcal{N}(\bm{z}_{lkt};\mumsg{\Psi_{0,lt}}{\bm{z}_{lkt}},\cmsg{\Psi_{0,lt}}{\bm{z}_{lkt}}),\\
	\msg{\Psi_{1,lkt}}{\bm{z}_{lkt}}&=\mathcal{N}(\bm{z}_{lkt};\mumsg{\Psi_{1,lkt}}{\bm{z}_{lkt}},\cmsg{\Psi_{1,lkt}}{\bm{z}_{lkt}}),\\
	\msg{\Psi_{1,lkt}}{x_{kt}}&=\pi_{1lkt}(x_{kt}),\\
	\msg{\Psi_{1,lkt}}{\bm{h}_{lk}} &= \mathcal{N}(\bm{h}_{lk},\mumsg{\Psi_{1,lkt}}{\bm{h}_{lk}},\cmsg{\Psi_{1,lkt}}{\bm{h}_{lk}}),\\
	\msg{\Psi_{2,lk}}{\bm{h}_{lk}}&=\mathcal{N}(\bm{h}_{lk},\mumsg{\Psi_{2,lk}}{\bm{h}_{lk}},\cmsg{\Psi_{2,lk}}{\bm{h}_{lk}}),\\
	\msg{\Psi_{3,kt}}{x_{kt}}&=\pi_{3kt}(x_{kt}).
\end{align*}
Please note that the dimensions of the parameters of all factor-to-variable messages are given by the type of distribution and the dimensions of the variable represented.
The variable-to-factor messages are just multiplications of factor-to-variable messages and are used to compute the update of the factor-to-variable messages.
Due to space limitation, we present the derivation only for some of the messages in the Appendix. Other messages that are not explicitly derived can be obtained by applying the rules presented in Section~\ref{EP_Graphs}.
\subsection{Initialization and Scheduling}
All messages with the exception of the constant messages representing a priori knowledge are initialized as uninformative distributions, i.e., for Gaussian messages the diagonal entries of the covariance matrices are set to infinity   and uniform \acp{PMF} are utilized for  categorical messages. In our implementation, all Gaussian messages are parameterized by the  precision matrix $\bm{\Lambda}=\bm{C}^{-1}$ and the mean vector $\bm{\gamma}=\bm{\Lambda}\bm{\mu}.$ Uninformative Gaussian messages are initialized by all-zero precision matrices. 

The scheduling is given in Algorithm \ref{alg1}. The first message to be computed in each iteration  is $\msg{\Psi_{1,lt}}{\bm{z}_{lkt}}.$ In the first iteration,  it is computed based on the \ac{CSI} from the Bayesian MMSE estimation together with the constant data prior to generate a first estimate of $z_{lkt}$. During this step, the initialized uninformed message $\msg{\Psi_{1,lt}}{\bm{z}_{lkt}}$ is updated to contain information. 
Note that the messages $\msg{\Psi_{2,lk}}{\bm{h}_{lk}}\equiv\Psi_{2,lk}$ and $\msg{\Psi_{3,kt}}{x_{kt}}\equiv\Psi_{3,kt}$  represent the a priori distributions on data and the Bayesian MMSE channel estimates, respectively, and are not modified by the message passing algorithm.
Hence, they are not included in the scheduling.

For our simulations, we used \emph{parallel} scheduling, i.e., the updates of the messages in the innermost for-loops are done independently suitable for efficient implementations.
\begin{algorithm} 
	\caption{bilinear-EP \ac{JCD} Scheduling} 
	\label{alg1} 
 \begin{algorithmic} 
		\REQUIRE Initialized Messages
		\FOR{number of iterations $I$}
		\FOR{all \acp{AP} $l$, \acp{UE} $k$ and symbol times $t$}
		\STATE Update 	$\msg{\Psi_{1,lkt}}{\bm{z}_{lkt}}$
		\ENDFOR
		\FOR{all \acp{AP} $l$, \acp{UE} $k$ and symbol times $t$}
		\STATE Update $\msg{\Psi_{0,lt}}{\bm{z}_{lkt}}$
		\ENDFOR
	    \FOR{all \acp{AP} $l$, \acp{UE} $k$ and symbol times $t$}
		\STATE Update $\msg{\Psi_{1,lkt}}{\bm{h}_{lk}}$ 
		\ENDFOR
	    \FOR{all \acp{AP} $l$, \acp{UE} $k$ and symbol times $t$}
		\STATE Update $	\msg{\Psi_{1,lkt}}{x_{kt}}$
		\ENDFOR
		\ENDFOR
	\end{algorithmic}
\end{algorithm}
\subsection{Update Rules}
In this section, we provide expressions to update  the  messages that appear in Algorithm~\ref{alg1}.
According to \eqref{eq:EP_fac_to_var_message}, the inputs to update a factor-to-variable message are always variable-to-factor messages. Each variable-to-factor message is computed by applying \eqref{eq:EP_var_to_fac_message}. As an example, we derive  the variable-to-factor message forwarded from variable node ${\bm{h}_{lk}}$ to factor node ${\Psi_{1,lkt}}$ as follows.
\begin{align*}
    \msg{\bm{h}_{lk}}{\Psi_{1,lkt}}&=\mathcal{N}(\bm{h}_{lk};\mumsg{\bm{h}_{lk}}{\Psi_{1,lkt}},\cmsg{\bm{h}_{lk}}{\Psi_{1,lkt}})\\
    &\propto\msg{\Psi_{2,lk}}{\bm{h}_{lk}}\prod_{\tilde{t}\neq t}\msg{\Psi_{1,lk\tilde{t}}}{\bm{h}_{lk}}.
\end{align*}
Since $\msg{\bm{h}_{lk}}{\Psi_{1,lkt}}$ is a Gaussian distribution, its computation reduces to determine its covariance and mean. By applying the Gaussian multiplication lemma\cite{Ngo2020}, we obtain  
\begin{align}
    \cmsg{\bm{h}_{lk}}{\Psi_{1,lkt}}&=\left(\cmsg{\Psi_{2,lk}}{\bm{h}_{lk}}^{-1}+\sum_{\tilde{t}\neq t} \cmsg{\Psi_{1,lk\tilde{t}}}{\bm{h}_{lk}}^{-1}\right)^{-1},
    \label{eq:C_h_Psi1}\\
    \begin{split}
    \mumsg{\bm{h}_{lk}}{\Psi_{1,lkt}}&=\cmsg{\bm{h}_{lk}}{\Psi_{1,lkt}}\bigg(\cmsg{\Psi_{2,lk}}{\bm{h}_{lk}}^{-1}\mumsg{\Psi_{2,lk}}{\bm{h}_{lk}}\\
    &+\sum_{\tilde{t}\neq t} \cmsg{\Psi_{1,lk\tilde{t}}}{\bm{h}_{lk}}^{-1}\mumsg{\Psi_{1,lk\tilde{t}}}{\bm{h}_{lk}}\bigg).
    \end{split}
    \label{eq:mu_h_Psi1}
\end{align}
  All other variable-to-factor messages can be computed analogously. Their derivation is omitted due to space constraints.
\\
In the following we provide the rules to  update  factor-to-variable messages based on the incoming variable-to-factor messages.

\noindent \textbf{Update of }$\msg{\Psi_{1,lkt}}{\bm{z}_{lkt}}=\mathcal{N}(\bm{z}_{lkt}; \mumsg{\Psi_{1,lkt}}{\bm{z}_{lkt}},  \cmsg{\Psi_{1,lkt}}{\bm{z}_{lkt}} ).$
The covariance and mean are given by 
\begin{align}
\cmsg{\Psi_{1,lkt}}{\bm{z}_{lkt}}\! &= \left(\bm{\Sigma}_{lkt}^{-1}-\cmsg{\bm{z}_{lkt}}{\Psi_{1,lkt}}^{-1}\right)^{-1},
\label{eq:C_Psi1_z}\\
\begin{split}
\mumsg{\Psi_{1,lkt}}{\bm{z}_{lkt}} &= \cmsg{\Psi_{1,lkt}}{\bm{z}_{lkt}}\Big(\bm{\Sigma}_{lkt}^{-1}\hat{\bm{z}}_{lkt}\\
&\quad-\cmsg{\bm{z}_{lkt}}{\Psi_{1,lkt}}^{-1}\mumsg{\bm{z}_{lkt}}{\Psi_{1,lkt}}\Big),
\end{split}
\label{eq:mu_Psi1_z}
\end{align}
where
\begin{align}
\hat{\bm{z}}_{lkt} &=\sum_{x'_{kt}\in\mathcal{S}}\omega(x'_{kt})\tilde{\bm{\mu}}(x'_{kt}),
\label{eq:z_hat}\\
\begin{split}
\bm{\Sigma}_{lkt} &= \sum_{x'_{kt}\in\mathcal{S}}	\omega(x'_{kt})\left(\tilde{\bm{\mu}}(x'_{kt})\tilde{\bm{\mu}}(x'_{kt})^H+\tilde{\bm{C}}(x'_{kt})\right)\\
&\quad-\bm{\hat{z}}_{lkt}\bm{\hat{z}}_{lkt}^H,
\end{split}
\label{eq:Sigma}\\
\omega(x_{kt}) &= \frac{\tilde{\omega}(x_{kt})}{\sum_{x'_{kt}\in\mathcal{S}}\tilde{\omega}(x'_{kt})},
\label{eq:omega}
\end{align}
with $\tilde{\omega}(x_{kt})$ defined in \eqref{eq:mult_messages_z}. Finally,
\begin{align}
\tilde{\bm{C}}(x_{kt})&=\left(\cmsg{\bm{z}_{lkt}}{\Psi_{1,lkt}}^{-1}+\frac{\cmsg{\bm{h}_{lk}}{\Psi_{1,lkt}}^{-1}}{\vert x_{kt}\vert^2}\right)^{-1},
\label{eq:C_tilde}\\
\begin{split}
\tilde{\bm{\mu}}(x_{kt})&=
\tilde{\bm{C}}(x_{kt})\Bigg(\cmsg{\bm{z}_{lkt}}{\Psi_{1,lkt}}^{-1}\mumsg{\bm{z}_{lkt}}{\Psi_{1,lkt}}\\
&\quad+\frac{\cmsg{\bm{h}_{lk}}{\Psi_{1,lkt}}^{-1}}{\vert x_{kt}\vert^2}x_{kt}\mumsg{\bm{h}_{lk}}{\Psi_{1,lkt}}\Bigg).
\end{split}
\label{eq:mu_tilde}
\end{align}

\noindent\textbf{Update of }$\msg{\Psi_{0,lt}}
{\bm{z}_{lkt}}=\mathcal{N}(\bm{z}_{lkt}; \mumsg{\Psi_{0,lt}}{\bm{z}_{lkt}},  \cmsg{\Psi_{0,lt}}{\bm{z}_{lkt}} ).$
The covariance and mean are given by 
\begin{align*}
    \cmsg{\Psi_{0,lt}}{\bm{z}_{lkt}}&=\sigma^2\bm{I}_N+\sum_{j\neq k} \cmsg{\Psi_{1,ljt}}{\bm{z}_{ljt}},\\
    \mumsg{\Psi_{0,lt}}{\bm{z}_{lkt}}&=\bm{y}_{lt}-\sum_{j\neq k}\mumsg{\Psi_{1,ljt}}{\bm{z}_{ljt}}.
\end{align*}

\noindent\textbf{Update of }$\msg{\Psi_{1,lkt}}{\bm{h}_{lk}}=\mathcal{N}(\bm{h}_{lk}; \mumsg{\Psi_{1,lkt}}{\bm{h}_{lt}},  \cmsg{\Psi_{1,lkt}}{\bm{h}_{lk}} ).$
The covariance and mean are given by 
\begin{align*}
\cmsg{\Psi_{1,lkt}}{\bm{h}_{lk}}\! &= \left(\hat{\bm{\Sigma}}_{lkt}^{-1}-\cmsg{\bm{h}_{lk}}{\Psi_{1,lkt}}^{-1}\right)^{-1},\\
\mumsg{\Psi_{1,lkt}}{\bm{h}_{lk}}\! &= \cmsg{\Psi_{1,lkt}}{\bm{h}_{lk}}\!\left(\hat{\bm{\Sigma}}_{lkt}^{-1}\hat{\bm{h}}_{lk}\!-\!\cmsg{\bm{h}_{lk}}{\Psi_{1,lkt}}^{-1}\mumsg{\bm{h}_{lk}}{\Psi_{1,lkt}}\!\right),
\end{align*}
where
\begin{align*}
\hat{\bm{h}}_{lk} &= \sum_{x'_{kt}\in\mathcal{S}}\omega(x'_{kt})\bar{\bm{\mu}}(x'_{kt}),\\
\hat{\bm{\Sigma}}_{lkt} &= \sum_{x'_{kt}\in\mathcal{S}}	\omega(x'_{kt})\left(\bar{\bm{\mu}}(x'_{kt})\bar{\bm{\mu}}(x'_{kt})^H+\bar{\bm{C}}(x'_{kt})\right)-\bm{\hat{h}}_{lk}\bm{\hat{h}}_{lk}^H.
\end{align*}
The factor $\omega(x_{kt})$ is defined in \eqref{eq:omega}. Finally,  \begin{align*}
\bar{\bm{C}}(x_{kt})&=\left(\cmsg{\bm{h}_{lk}}{\Psi_{1,lkt}}^{-1}+\vert x_{kt}\vert^2\cmsg{\bm{z}_{lkt}}{\Psi_{1,lkt}}^{-1}\right)^{-1},\\
\bar{\bm{\mu}}(x_{kt})&=
\bar{\bm{C}}(x_{kt})\Bigg(\vert x_{kt}\vert^2\frac{\cmsg{\bm{z}_{lkt}}{\Psi_{1,lkt}}^{-1}}{x_{kt}}\mumsg{\bm{z}_{lkt}}{\Psi_{1,lkt}}\\
&\quad+\cmsg{\bm{h}_{lk}}{\Psi_{1,lkt}}^{-1}\mumsg{\bm{h}_{lk}}{\Psi_{1,lkt}}\Bigg).
\end{align*}

\noindent\textbf{Update of }$\msg{\Psi_{1,lkt}}{\bm{x}_{kt}}=\pi_{1,lkt}(x_{kt}).$
The probabilities of the categorical distribution are given by
\begin{equation}
\msg{\Psi_{1,lkt}}{x_{kt}}=\frac{\tilde{\gamma}(x_{kt})}{\sum_{x'_{kt}\in\mathcal{S}}\tilde{\gamma}(x'_{kt})}
\label{eq:m_Ps1_x}
\end{equation}
with
\begin{align*}
\tilde{\gamma}(x_{kt}):=\mathcal{N}(\bm{0};\mumsg{\bm{z}_{lkt}}{\Psi_{1,lkt}}-x_{kt}\mumsg{\bm{h}_{lk}}{\Psi_{1,lkt}},\\
\cmsg{\bm{z}_{lkt}}{\Psi_{1,lkt}}+\vert x_{kt}\vert^2\cmsg{\bm{h}_{lk}}{\Psi_{1,lkt}}).
\end{align*}

\subsection{Inference}
The desired estimates can be obtained computing the distributions  $\hat{p}_{x_{kt}}$ and $\hat{p}_{\bm{h}_{lk}}$ as follows
\begin{equation*}
\hat{x}_{kt}=\underset{x_{kt}\in\mathcal{S}}{\arg\max}\,\hat{p}_{x_{kt}}=\underset{x_{kt}\in\mathcal{S}}{\arg\max}\,\msg{\Psi_{3,kt}}{x_{kt}}\prod_{l=1}^L\msg{\Psi_{1,lkt}}{x_{kt}}
\end{equation*}
and
\begin{align*}
	\hat{\bm{h}}_{lk}=\underset{\bm{h}_{lk}\in\mathbb{C}^N}{\arg\max}\,\hat{p}_{\bm{h}_{lk}}&=\underset{\bm{h}_{lk}\in\mathbb{C}^N}{\arg\max}\,\msg{\Psi_{2,lk}}{\bm{h}_{lk}}\prod_{t=1}^T\msg{\Psi_{1,lkt}}{\bm{h}_{lk}}\\
 &=\underset{\bm{h}_{lk}\in\mathbb{C}^N}{\arg\max}\,\mathcal{N}(\bm{h}_{lk};\bm{\mu}_{\mathrm{tot};\bm{h}},\bm{C}_{\mathrm{tot};\bm{h}})\\
 &=\bm{\mu}_{\mathrm{tot};\bm{h}},
\end{align*}
where \begin{align*}
\bm{C}_{\mathrm{tot};\bm{h}}&=\left(\cmsg{\Psi_{2,lk}}{\bm{h}_{lk}}^{-1}+\sum_{t=1}^T\cmsg{\Psi_{1,lkt}}{\bm{h}_{lk}}^{-1}\right)^{-1},\\
\bm{\mu}_{\mathrm{tot};\bm{h}}&=\bm{C}_{\mathrm{tot};\bm{h}}\Bigg(\cmsg{\Psi_{2,lk}}{\bm{h}_{lk}}^{-1}\mumsg{\Psi_{2,lk}}{\bm{h}_{lk}}\\
&\quad+\sum_{t=1}^T\cmsg{\Psi_{1,lkt}}{\bm{h}_{lk}}^{-1}\mumsg{\Psi_{1,lkt}}{\bm{h}_{lk}}\Bigg)
\end{align*}
are obtained by applying the Gaussian multiplication lemma~\cite{Ngo2020}.
Interestingly, the availability of the posterior distributions allows us to determine the reliability of the estimates $\hat{x}_{kt}$ and  $\hat{\bm{h}}_{lk}.$ 

\subsection{Algorithm Stability}
 Each parameter of a distribution is iteratively computed by a \textit{soft update}\cite{Cespedes2014}, i.e., given the old parameters $\bm{\theta}^{(i-1)}$, the new parameters $\bm{\theta}^{(i)}$ are computed according to the message-passing rules  and then updated as
\begin{equation*}
    \bm{\theta}^{(i)}\gets \eta\,\bm{\theta}^{(i)}+(1-\eta)\,\bm{\theta}^{(i-1)},
\end{equation*}
where $ \eta \in [0,1]$  is the soft update parameter.
Additionally,  the computation of some messages, e.g., \eqref{eq:C_Psi1_z} may lead to precision matrices with negative eigenvalues. Inspired by~\cite{Ngo2020}, we set to zero the negative eigenvalues to assure positive semi-definiteness as required for Gaussian distributions.
\subsection{Analysis of Fronthaul Load}
In this section we discuss a distributed implementation of the bilinear-EP \ac{JCD} with special attention to the messages exchanged through the fronthaul. 
As apparent from  \eqref{eq:constituents_message_x_kt}  in the Appendix,  \ac{AP} $l$ requires the messages  $\msg{\Psi_{1,\tilde{l}kt}}{x_{kt}}$  from the other \acp{AP} $\tilde{l}\neq l$. The data prior $\msg{\Psi_{3,kt}}{x_{kt}}\equiv\Psi_{3,kt}$ is assumed to be uninformative and thus not needed.
The messages are exchanged via the \ac{CPU} which computes, based on the incoming messages from all \acp{AP}, the following message 
\begin{equation}
    m_{x_{kt};\mathrm{tot}}\propto\prod_{l=1}^L \msg{\Psi_{1,lkt}}{x_{kt}},\label{eq:m_x_kt_tot_def}
\end{equation}
and then forwards them to the \acp{AP}.
Then,  \ac{AP} $l$ can remove its own message from the incoming  message  from the \ac{CPU} to obtain the desired message as in  \eqref{eq:constituents_message_x_kt} , i.e.,
\begin{equation*}
    \msg{x_{kt}}{\Psi_{1,lkt}} \propto\frac{m_{x_{kt};\mathrm{tot}}}{\msg{\Psi_{1,lkt}}{x_{kt}}}.
\end{equation*}

In the following, we quantify the fronthaul load for the communication between \acp{AP} and \ac{CPU} in terms of number of messages  per iteration. 
Each \ac{AP}  transmits $KT$ messages $\msg{\Psi_{1,lkt}}{x_{kt}}$ to the \ac{CPU}. Additionally, the message $m_{x_{kt};\mathrm{tot}}$ is transmitted to each \ac{AP} totalling $LKT$ messages to be transmitted.

\subsection{Computational Complexity}\label{Complexity}
The order of  computational  complexity at the \acp{AP} is mainly determined  by the computation of inverse matrices  of  dimension $N$. Then, we focus on the messages which require matrix inversion and present the  highest computational complexity, namely, messages $\msg{\Psi_{1,lkt}}{\bm{z}_{lkt}}$ and $\msg{\Psi_{1,lkt}}{\bm{h}_{lk}}$. Their order of complexity can be obtained from  \eqref{eq:C_tilde} and \eqref{eq:mult_messages_z}. We can observe that  $KT|\mathcal{A}|$ weighted sums of covariance matrices per \ac{AP} need to be inverted  where $\mathcal{A}:=\{a\in\mathbb{R}^+\,|\,a=|x|^2,x\in\mathcal{S}\}$ denotes the set  of distinct amplitudes in the signal constellation set $\mathcal{S}$. Therefore, at the \ac{AP}, the complexity  order  per iteration is $\mathcal{O}(KT|\mathcal{A}|N^3)$. 
The computation of the message $m_{x_{kt};\mathrm{tot}}$ in \eqref{eq:m_x_kt_tot_def} at the \ac{CPU} requires the multiplication of real-valued scalars of order $\mathcal{O}(LKT|\mathcal{S}|)$.
Therefore, the order of the computational  complexity at the CPU is $\mathcal{O}(LKT|\mathcal{S}|)$ per iteration of the bilinear-EP \ac{JCD} algorithm. The total computational complexity that takes into account the processing in each \acs{AP} and at the \acs{CPU} is then given by $\mathcal{O}\left(LKT(|\mathcal{S}|+|\mathcal{A}|N^3)\right)$.
\section{Simulation Results}\label{sim_res}
\subsection{Setting}

In this section we present the setting that we utilized for our simulation.
We consider a square surface with 400m side length and position $L=16$ \acp{AP} on a rectangular grid, i.e., on the  points of set $\{(i\times \frac{400}{3}\mathrm{m},j\times \frac{400}{3}\mathrm{m})\,|\,i,j\in
\{0,1,2,3\}\}$ as in \cite{Bjoernson2020}. Each \ac{AP} is equipped with $N=1$ antenna. The $K=8$ \acp{UE} are uniformly randomly distributed over the square surface. Each \ac{UE} transmits  with a  power of $p = 14\,\mathrm{dBm}$  and at each \ac{AP} the received signal is impaired by an  additive  white noise with variance  $v = -96\,\mathrm{dBm}$ .
The diagonal elements of the channel covariance matrix $\bm{\Xi}_l$ are determined by the distance between the \ac{UE} $k$ and \ac{AP} $l$ using the fading model in~\cite{Bjoernson2020}, i.e., for \ac{UE} $k$ and for all co-located antennas $n$ at \ac{AP}~$l$
\begin{equation*}
    [\bm{\Xi}_l]_{nk,nk}[\mathrm{dB}] = -30.5-36.7\log_{10}\left(\frac{d_{kl}}{1m}\right),
\end{equation*}
where $d_{kl}$ is the distance between \ac{UE} $k$ and \ac{AP} $l$. 
The pilot sequences are orthogonal, specifically,  $\bm{X}_p=\bm{I}$, $P=K.$ The variability of the scenario is captured by sampling 300 realizations of the positions of the \acp{UE}. In each position we perform \SI{e4} transmissions  with different  small-scale fading realizations. In each transmission, $T\in\{10,100\}$ symbols are sent per \ac{UE}.    We use 4-QAM as modulation scheme. Our numerical results  are obtained with $I=10$ iterations and we use $\eta=0.7$ as soft update parameter.

\subsection{Performance evaluation and complexity analysis}
\begin{table*}[tbp]
\normalsize
\caption{Computational complexity of distributed and centralized algorithms}
\centering
	\begin{tabular}{ |p{3cm}||p{4cm}|p{3cm}|p{5.8cm}|  }
 \hline
 Algorithm & Complexity at \ac{AP} &Complexity at \ac{CPU} &Total Complexity\\
 \hline
 bilinear-EP JCD   & $\mathcal{O}(KT|\mathcal{A}|N^3)$    &$\mathcal{O}(LKT|\mathcal{S}|)$&  $\mathcal{O}\left(LKT(|\mathcal{S}|+|\mathcal{A}|N^3)\right)$\\
 He et al. ICD & $\mathcal{O}\left(TKN^2+(T+P)^3N^3\right)$ & $\mathcal{O}(TK^2)$ & $\mathcal{O}\left(L(TKN^2+(T+P)^3N^3)+TK^2\right)$\\
 Ngo et al.  & - & - & $\mathcal{O}\left(K^6|\mathcal{S}|^T)\right)$\\
 Centralized MMSE & - & - & $\mathcal{O}\left(K\left(LN^3+K^2+LNT+|\mathcal{S}|T\right)\right)$\\
 \hline
\end{tabular}
\label{tab:Computational Complexity}

\end{table*}
In this section we present our simulation results and discuss the performance-complexity trade-off. As usual in the investigation of \ac{CF-MaMIMO} to analyze the \ac{QoS} distribution, we study the performance of the proposed algorithm in terms of the  empirical \acp{CDF} of the \acl{SER}  $\mathrm{SER}:=\estemp{\mathbbm{1}_{\hat{x}\neq x}}$ for detection and \ac{NMSE}  defined as $\mathrm{NMSE}:=\estemp{\frac{\Vert \hat{\bm{h}}-\bm{h}\Vert^2}{\Vert \bm{h}\Vert^2}}$ for channel estimation.

The samples for the empirical \acp{CDF} of the \ac{SER} and \ac{NMSE} are obtained per each large-scale channel fading and \ac{UE} realization and per each large-scale channel fading, \ac{AP} and \ac{UE} realization, respectively, by averaging the  instantaneous values corresponding to different small-scale realizations. Additionally, in the \ac{CDF} of the \ac{NMSE} we omit  weak channels, i.e., channels where the received power is lower than the noise variance to avoid to take into account errors on weak and insignificant channels.
In our simulation, we consider  four baseline schemes, namely, the detector in \cite{He2021} assuming perfect \ac{CSI}, the \ac{ICD} algorithm in \cite{He2023}, which is also a semi-blind algorithm based on \ac{EP} with polynomial complexity  order, a modified version of our proposed bilinear-\ac{EP} \ac{JCD} algorithm with perfect channel knowledge which provides a lower bound to the \ac{SER}, and a centralized conventional receiver based on a Bayesian MMSE channel estimation algorithm and a subsequent linear MMSE filter for data detection.

Next, we compare the computational complexity order of the proposed bilinear-EP algorithm with the baseline EP algorithm in \cite{He2023}, a non-coherent centralized EP algorithm proposed in \cite{Ngo2020}, and the baseline centralized approach based on linear MMSE filters for both channel estimation and detection.
 The results are summarized in Table \ref{tab:Computational Complexity}.  
 Notably, the proposed bilinear-EP JCD algorithm and the algorithm in  \cite{He2023} have linear and cubic (quadratic) complexity order in $T(K)$, respectively. Thus, the proposed algorithm is more efficient for long data sequences and a large number of UEs. However in our approach, the highest order term $N^3$ scales linearly with the number of UEs $K$. As the bilinear-EP JCD algorithm, the non-coherent centralized EP algorithm in  \cite{Ngo2020} adopts an exact categorical distribution for data.  When it is used for uncoded transmission, as in the proposed approach, its complexity order is given by $\mathcal{O}\left(K^6|\mathcal{S}|^T)\right)$. Due to its exponential complexity in the length of the symbol sequence length $T$, the required computational efforts are  prohibitive for a comparative study of the algorithm performance. Essentially, the algorithm in  \cite{Ngo2020}  approximates a \ac{MAP} decoder whereas our algorithm approximates a \ac{MAP} detector. It is worth noting that the extension of proposed bilinear-EP JCD algorithm with a decoder is straightforward through the use of loopy belief propagation which is known to have polynomial complexity as well. In contrast to the other algorithms, the conventional centralized MMSE algorithm is not iterative in nature and, thus, a fair comparison with the other receivers which require multiple iterations is not straightforward. 

\begin{figure}
	\includegraphics[width=\linewidth]{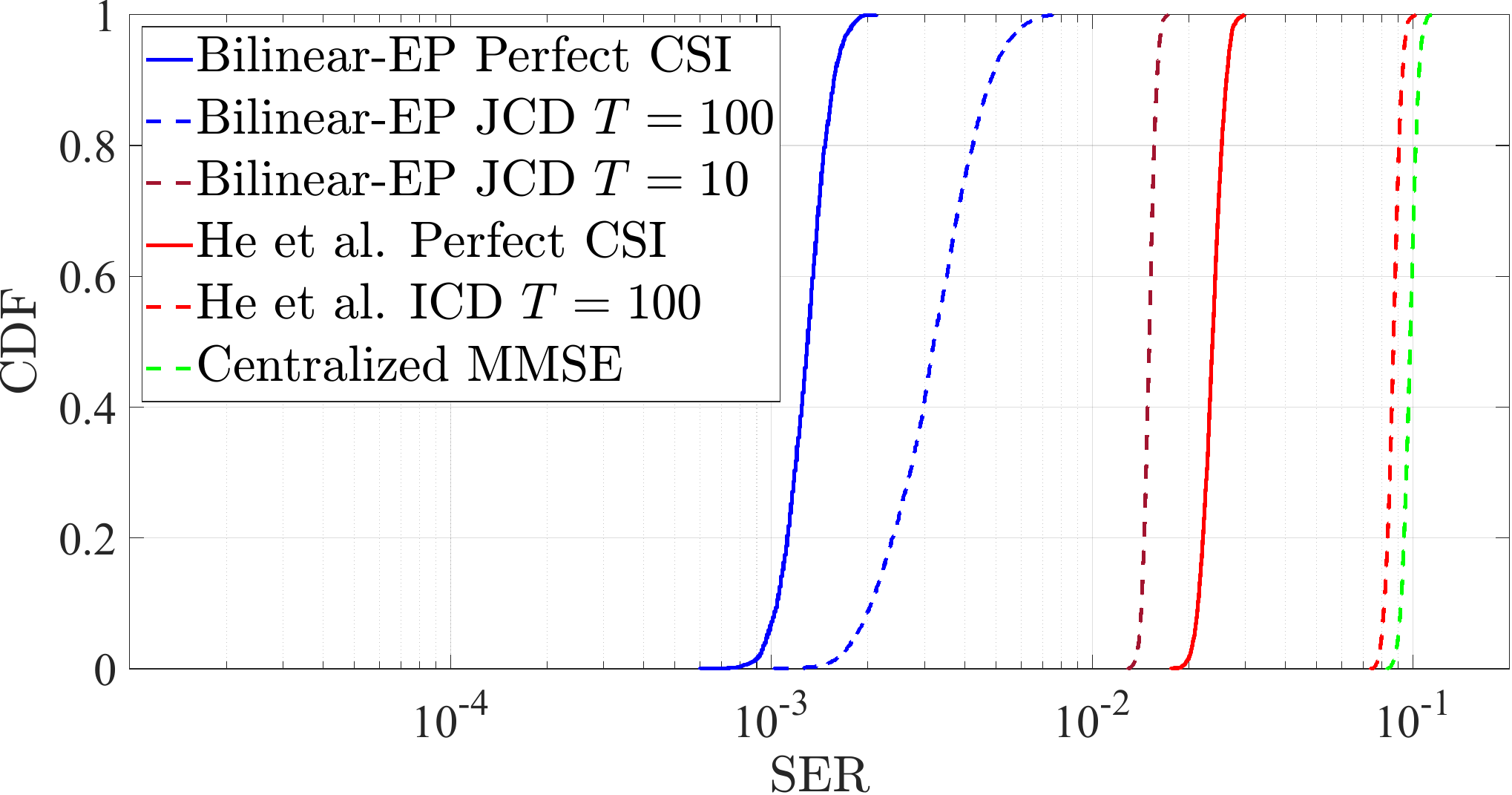}
	\caption{\ac{CDF} of \ac{SER} for $L=16,K=8,N=1$.}
	\label{fig:SER_CDF}
\end{figure}

\begin{figure}
		\includegraphics[width=\linewidth]{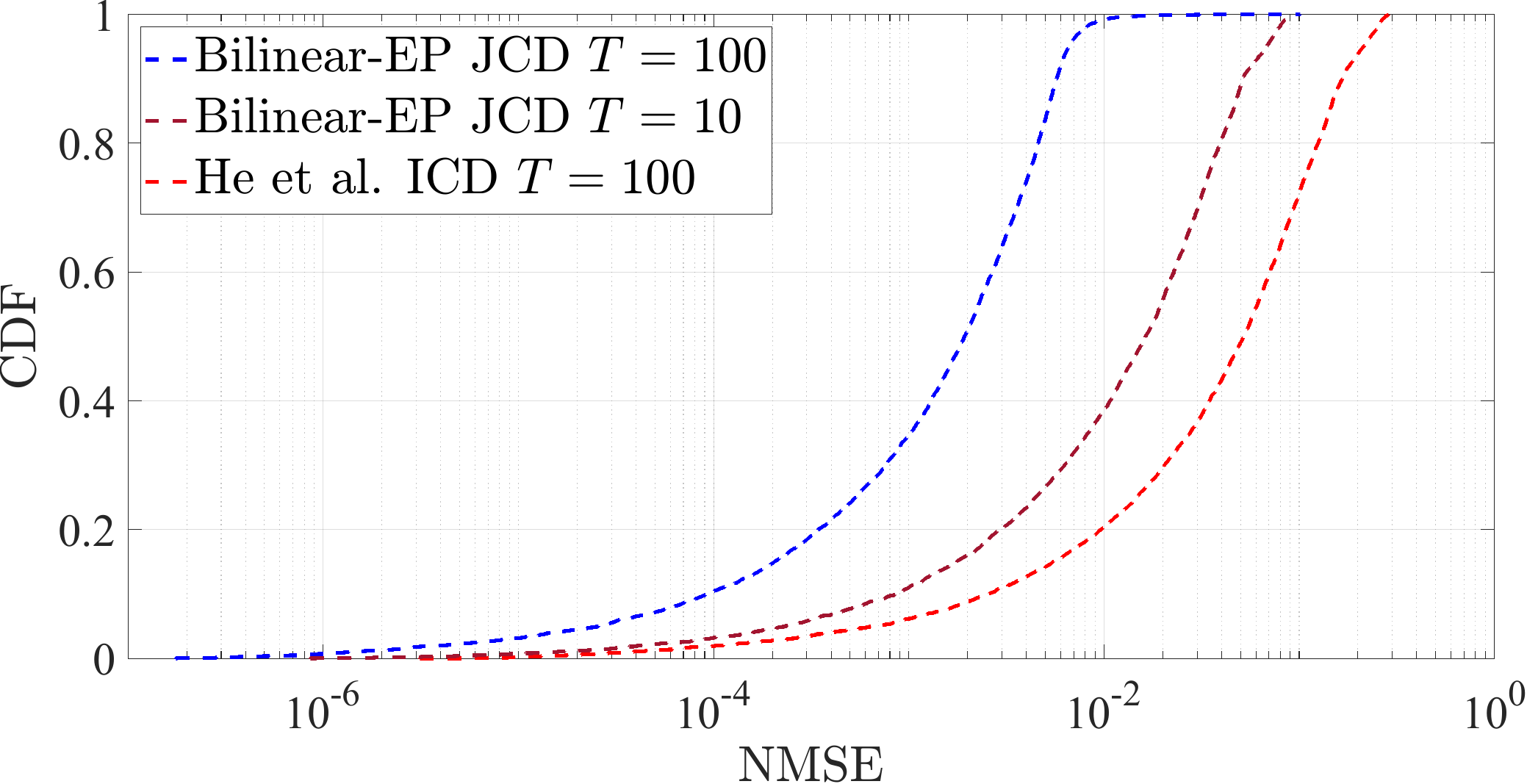}
		\caption{\ac{CDF} of \ac{NMSE} for $L=16,K=8,N=1$.}
		\label{fig:NMSE_CDF}
\end{figure}

In the following, we analyze the performance.
Fig. \ref{fig:SER_CDF}  shows the empirical \ac{CDF} of the \ac{SER} of the proposed bilinear-\ac{EP} \ac{JCD} for $T=\{10, 100\}$ and the above mentioned baselines.  For both values of $T$ the proposed bilinear-EP algorithm outperforms the \ac{ICD} algorithm in \cite{He2023} by more than one order of magnitude, while the complexity of bilinear-EP is only linear in $T$ and not cubic. Our approach also outperforms the \ac{EP}-based detector in \cite{He2021} with perfect \ac{CSI} and the centralized MMSE approach. The gap in performance with the second baseline approach indicates that the Gaussian approximation of the data symbol distribution  and the averaging of the messages in \cite{He2021, He2023} determine a degradation of performance. We observe  an improvement of the \ac{CDF} of the  \ac{SER} obtained with our bilinear-\ac{EP} \ac{JCD} algorithm when  the length of the transmitted data symbols increases  from $T=10$ to $T=100$. The performance of our modified algorithm obtained assuming  perfect \ac{CSI} shows that there is room for  further improvement.
 Fig. \ref{fig:NMSE_CDF} shows the \ac{CDF} of the \ac{NMSE}. We observe an improvement of the \ac{NMSE} by more than an order of magnitude of our proposed algorithm compared to the \ac{ICD} scheme in \cite{He2023}. Additionally, the performance of our channel estimation  also improves when $T$ increases.


\begin{figure*}[tbp]
\normalsize
\begin{align}
    \begin{split}
	&\msg{\bm{h}_{lk}}{\Psi_{1,lkt}}\left(\frac{\bm{z}_{lkt}}{x_{kt}}\right)=\frac{1}{\pi^N\mathrm{det}\left(\cmsg{\bm{h}_{lk}}{\Psi_{1,lkt}}\right)}\cdot\exp\left(-\left(\frac{\bm{z}_{lkt}}{x_{kt}}-\mumsg{\bm{h}_{lk}}{\Psi_{1,lkt}}\right)^H\cmsg{\bm{h}_{lk}}{\Psi_{1,lkt}}^{-1}\left(\frac{\bm{z}_{lkt}}{x_{kt}}-\mumsg{\bm{h}_{lk}}{\Psi_{1,lkt}}\right)\right)\\
	&=|x_{kt}|^{2N}\mathcal{N}(\bm{z}_{lkt};x_{kt}\mumsg{\bm{h}_{lk}}{\Psi_{1,lkt}},\vert x_{kt} \vert^2\cmsg{\bm{h}_{lk}}{\Psi_{1,lkt}})
    \end{split}
    \label{Gaussian Transformation}\tag{29}\\
\begin{split}
&\msg{\bm{z}_{lkt}}{\Psi_{1,lkt}}(\bm{z}_{lkt})\msg{x_{kt}}{\Psi_{1,lkt}}(x_{kt})\mathcal{N}(\bm{z}_{lkt};x_{kt}\mumsg{\bm{h}_{lk}}{\Psi_{1,lkt}},\vert x_{kt} \vert^2\cmsg{\bm{h}_{lk}}{\Psi_{1,lkt}})\\
&=\mathcal{N}(\bm{z}_{lkt};\tilde{\bm{\mu}}(x_{kt}),\tilde{\bm{C}}(x_{kt}))
\cdot\underbrace{\msg{x_{kt}}{\Psi_{1,lkt}}(x_{kt})\mathcal{N}(\bm{0};\mumsg{\bm{z}_{lkt}}{\Psi_{1,lkt}}\!-x_{kt}\mumsg{\bm{h}_{lk}}{\Psi_{1,lkt}},\cmsg{\bm{z}_{lkt}}{\Psi_{1,lkt}}+\vert x_{kt}\vert^2\cmsg{\bm{h}_{lk}}{\Psi_{1,lkt}})}_{:=\tilde{\omega}(x_{kt})}
\end{split}
\label{eq:mult_messages_z}\tag{30}
\end{align}

\hrulefill
\end{figure*}

\section{Conclusion}\label{concl}
In this paper, we considered a \ac{CF-MaMIMO} system  and tackled  the quest of low-complexity \ac{JCD} with near-optimal performance and robustness  to pilot contamination. We derived a blind or semi-blind  distributed \ac{JCD} algorithm by formulating the problem in the framework of bilinear inference and obtaining the solution as an unfolding of a message passing algorithm incorporating \ac{EP}-rules over a factor graph. 
The appealing features of the proposed algorithm stem from our choice of the approximate  posterior joint distribution of data symbols and channels.  Our simulation results show that the
proposed scheme significantly outperforms the selected baseline schemes based on the detector in \cite{He2021},  which assumes perfect \ac{CSI}, and the \ac{ICD} algorithm in \cite{He2023}. Additionally,  bilinear-EP \ac{JCD} has polynomial computational complexity and allows for straightforward embedding of state-of-the art \ac{SISO} decoders, such that the decoding complexity can also be kept of polynomial order. This enables further investigations on the rate that can be achieved using our proposed JCD algorithm.

\section*{Appendix}
In this section, we derive some of the factor-to-variable messages. When it does not cause ambiguity, we adopt the following abbreviated notation $\msg{\Psi_{1,lkt}}{\bm{z}_{lkt}}=\msg{\Psi_{1,lkt}}{\bm{z}_{lkt}}(\bm{z}_{lkt})$.
\vspace{0.5em}\\
\textbf{Derivation of message} $\msg{\Psi_{1,lkt}}{x_{kt}}$\\
Message $\msg{\Psi_{1,lkt}}{x_{kt}}$ is obtained by applying  \eqref{eq:EP_fac_to_var_message} and \eqref{eq:message_projection}. This computation requires the knowledge of messages from variable nodes to factor nodes, namely,  $\msg{\bm{h}_{lk}}{\Psi_{1,lkt}},$ $\msg{\bm{z}_{lkt}}{\Psi_{1,lkt}},$ and $ \msg{x_{kt}}{\Psi_{1,lkt}}$ as we can evince from the factor graph in Fig. \ref{fig:Delta_JCD_L_AP}.  In the following, we derive these messages by applying (\ref{eq:EP_var_to_fac_message}). 
Thus, we obtain
$\msg{\bm{h}_{lk}}{\Psi_{1,lkt}}(\bm{h}_{lk})\propto \msg{\Psi_{2,lk}}{\bm{h}_{lk}}\prod_{\tilde{t}\neq t}\msg{\Psi_{1,lk\tilde{t}}}{\bm{h}_{lk}}$ which is a Gaussian distribution with covariance matrix and mean given by \eqref{eq:C_h_Psi1} and \eqref{eq:mu_h_Psi1}, respectively.
As product of Gaussian distributions, the mean and covariance matrix above are obtained by applying the Gaussian multiplication lemma, see, e.g., \cite{Ngo2020}. 
Similarly, we derive the other variable-to-factor messages as $\msg{\bm{z}_{lkt}}{\Psi_{1,lkt}}=\msg{\Psi_{0,lt}}{\bm{z}_{lkt}}$ and 
\begin{equation}
\msg{x_{kt}}{\Psi_{1,lkt}}\propto\msg{\Psi_{3,kt}}{x_{kt}}\prod_{\tilde{l}\neq l} \msg{\Psi_{1,\tilde{l}kt}}{x_{kt}}
\label{eq:constituents_message_x_kt}
\end{equation}
which is a categorical distribution.
We normalize it as follows
\begin{align}
    &\msg{x_{kt}}{\Psi_{1,lkt}}(x_{kt})=\nonumber\\
    &\frac{\msg{\Psi_{3,kt}}{x_{kt}}(x_{kt}) \prod_{\tilde{l}\neq l} \msg{\Psi_{1,\tilde{l}kt}}{x_{kt}}(x_{kt})}{\sum_{x_{kt}\in\mathcal{S}}\msg{\Psi_{3,kt}}{x_{kt}}(x_{kt}) \prod_{\tilde{l}\neq l} \msg{\Psi_{1,\tilde{l}kt}}{x_{kt}}(x_{kt})}.
\end{align}
By utilizing the variable-to-factor messages impinging on the factor node $\Psi_{1,lkt}$ computed above, we apply~\eqref{eq:message_projection} to obtain the distribution before projection as follows
\begin{align}
\qmsg{\Psi_{1,lkt}}{x_{kt}}&\propto\int \delta(\bm{z}_{lkt}-x_{kt}\bm{h}_{lk}) \nonumber \\
&\times\msg{\bm{h}_{lk}}{\Psi_{1,lkt}}\msg{\bm{z}_{lkt}}{\Psi_{1,lkt}}d\bm{z}_{lkt}d\bm{h}_{lk}\msg{x_{kt}}{\Psi_{1,lkt}}. \nonumber \\
&\propto\int\frac{1}{\vert x_{kt}\vert^{2N}}\msg{\bm{h}_{lk}}{\Psi_{1,lkt}}\left(\frac{\bm{z}_{lkt}}{x_{kt}}\right) \nonumber \\
&\times\msg{\bm{z}_{lkt}}{\Psi_{1,lkt}}(\bm{z}_{lkt})d\bm{z}_{lkt}\msg{x_{kt}}{\Psi_{1,lkt}}, \label{eq:distribution_q_psi_1_x}
\end{align}
where the last expression is derived by applying the sifting property of the Dirac delta function, see, e.g., \cite{Candan2021}.  Then, the factor $\msg{\bm{h}_{lk}}{\Psi_{1,lkt}}\left(\frac{\bm{z}_{lkt}}{x_{kt}}\right)$ in (\ref{eq:distribution_q_psi_1_x})   can be rewritten  as shown in \eqref{Gaussian Transformation}\setcounter{equation}{29} at the top of the page.  Therefore, using the Gaussian multiplication lemma we obtain
\begin{align*}
\qmsg{\Psi_{1,lkt}}{x_{kt}}&\propto\mathcal{N}(\bm{0};\mumsg{\bm{z}_{lkt}}{\Psi_{1,lkt}}-x_{kt}\mumsg{\bm{h}_{lk}}{\Psi_{1,lkt}},\cmsg{\bm{z}_{lkt}}{\Psi_{1,lkt}}\\
&\quad+\vert x_{kt}\vert^2\cmsg{\bm{h}_{lk}}{\Psi_{1,lkt}})\cdot\msg{x_{kt}}{\Psi_{1,lkt}}.
\end{align*}
Next, we observe that  $\qmsg{\Psi_{1,lkt}}{x_{kt}}$ is already a categorical distribution in the exponential family. Then, the projection operator in  (\ref{eq:EP_fac_to_var_message})  leaves its argument unchanged, i.e.,  $\mathrm{proj}\{\qmsg{\Psi_{1,lkt}}{x_{kt}}\}=\qmsg{\Psi_{1,lkt}}{x_{kt}}$.  
Finally, by applying \eqref{eq:EP_fac_to_var_message}, we obtain the message  $\msg{\Psi_{1,lkt}}{x_{kt}}$  which is given in \eqref{eq:m_Ps1_x}.
\vspace{0.5em}\\
\textbf{Derivation of message} $\msg{\Psi_{1,lkt}}{\bm{z}_{lkt}}$\\
For the derivation of  message $\msg{\Psi_{1,lkt}}{\bm{z}_{lkt}}$ we consider again factor node $\Psi_{1,lkt}$ in Fig. \ref{fig:Delta_JCD_L_AP}. Thus, the same  variable-to-factor messages  computed above are necessary to determine $\qmsg{\Psi_{1,lkt}}{\bm{z}_{lkt}}$.
Analogously, the distribution before  projection onto the family of exponential functions  is
\begin{align*}
\qmsg{\Psi_{1,lkt}}{\bm{z}_{lkt}}&\propto\sum_{x'_{kt}\in\mathcal{S}}\msg{\bm{z}_{lkt}}{\Psi_{1,lkt}}(\bm{z}_{lkt})\frac{1}{\vert x'_{kt}\vert^{2N}}\\
&\quad\cdot\msg{\bm{h}_{lk}}{\Psi_{1,lkt}}\left(\frac{\bm{z}_{lkt}}{x'_{kt}}\right)\msg{x_{kt}}{\Psi_{1,lkt}}.
\end{align*}
We observe that $\qmsg{\Psi_{1,lkt}}{\bm{z}_{lkt}}$ is a Gaussian mixture in $|\mathcal{S}|$ components with the parameters $\tilde{\bm{C}}(x_{kt})$ and $\tilde{\bm{\mu}}(x_{kt})$ which depend on the specific value $x_{kt}$ and are defined in \eqref{eq:C_tilde} and \eqref{eq:mu_tilde}, respectively, which is obtained by applying again \eqref{Gaussian Transformation} and then the  Gaussian multiplication lemma. Additionally, each component of the Gaussian mixture distribution  is weighted by the unnormalized factor $\tilde{\omega}(x_{kt})$ given in \eqref{eq:mult_messages_z}\setcounter{equation}{30} at the top of the page.
Then, as in (\ref{eq:message_projection}) we have to project the Gaussian mixture distribution onto the family of Gaussian distributions, i.e., we need to determine the Gaussian distribution $\mathcal{N}(\bm{z}_{lkt};\hat{\bm{z}}_{lkt},\bm{\Sigma}_{lkt}):=\mathrm{proj}\{\qmsg{\Psi_{1,lkt}}{\bm{z}_{lkt}}\}$
whose moments are matched to the distribution $\qmsg{\Psi_{1,lkt}}{\bm{z}_{lkt}}$. Denoting with $\omega(x_{kt})$ the normalized weights obtained from $\tilde{\omega}(x_{kt})$ given in \eqref{eq:omega}, the parameters $\hat{\bm{z}}_{lkt}$ and $\bm{\Sigma}_{lkt}$ of the moment matched distribution are shown in \eqref{eq:z_hat} and \eqref{eq:Sigma}, respectively.
Finally, we have
\begin{equation*}
\msg{\Psi_{1,lkt}}{\bm{z}_{lkt}}\propto\frac{\mathrm{proj}\{\qmsg{\Psi_{1,lkt}}{\bm{z}_{lkt}}\}}{\msg{\bm{z}_{lkt}}{\Psi_{1,lkt}}}=\frac{\mathcal{N}(\bm{z}_{lkt};\hat{\bm{z}}_{lkt},\bm{\Sigma}_{lkt})}{\msg{\bm{z}_{lkt}}{\Psi_{1,lkt}}}.
\end{equation*}
The parameters $\cmsg{\Psi_{1,lkt}}{\bm{z}_{lkt}}$ and $\mumsg{\Psi_{1,lkt}}{\bm{z}_{lkt}}$ of the updated message are then given by the Gaussian multiplication lemma shown in \eqref{eq:C_Psi1_z} and \eqref{eq:mu_Psi1_z}, respectively.\vspace{0.5em}

\bibliographystyle{ieeetr}
\bibliography{references}

\begin{thebibliography}{10}

\bibitem{Ngo2017}
H.~Q. Ngo, A.~Ashikhmin, H.~Yang, E.~G. Larsson, and T.~L. Marzetta, ``Cell-free massive {MIMO} versus small cells,'' {\em IEEE Transactions on Wireless Communications}, vol.~16, no.~3, pp.~1834--1850, 2017.

\bibitem{Ngo2018}
H.~Q. Ngo, L.-N. Tran, T.~Q. Duong, M.~Matthaiou, and E.~G. Larsson, ``On the total energy efficiency of cell-free massive {MIMO},'' {\em IEEE Transactions on Green Communications and Networking}, vol.~2, no.~1, pp.~25--39, 2018.

\bibitem{Yang2018}
H.~Yang and T.~L. Marzetta, ``Energy efficiency of massive {MIMO}: Cell-free vs. cellular,'' in {\em Proc. of IEEE 87th Vehicular Technology Conference (VTC Spring)}, pp.~1--5, 2018.

\bibitem{Ammar2022}
H.~A. Ammar, R.~Adve, S.~Shahbazpanahi, G.~Boudreau, and K.~V. Srinivas, ``User-centric cell-free massive {MIMO} networks: A survey of opportunities, challenges and solutions,'' {\em IEEE Communications Surveys {\&} Tutorials}, vol.~24, no.~1, pp.~611--652, 2022.

\bibitem{Yin2014}
H.~Yin, D.~Gesbert, and L.~Cottatellucci, ``Dealing with interference in distributed large-scale {MIMO} systems: A statistical approach,'' {\em IEEE Journal of Selected Topics in Signal Processing}, vol.~8, no.~5, pp.~942--953, 2014.

\bibitem{Chen2018}
Z.~Chen and E.~Bj{\"o}rnson, ``Channel hardening and favorable propagation in cell-free massive {MIMO} with stochastic geometry,'' {\em IEEE Transactions on Communications}, vol.~66, no.~11, pp.~5205--5219, 2018.

\bibitem{Gholami2020a}
R.~Gholami, L.~Cottatellucci, and D.~Slock, ``Favorable propagation and linear multiuser detection for distributed antenna systems,'' in {\em Proc. of IEEE International Conference on Acoustics, Speech and Signal Processing (ICASSP)}, 2020.

\bibitem{Gholami2020b}
R.~Gholami, L.~Cottatellucci, and D.~Slock, ``Channel models, favorable propagation and {MultiStage} linear detection in cell-free massive {MIMO},'' in {\em Proc. of IEEE International Symposium on Information Theory (ISIT)}, 2020.

\bibitem{Marzetta2010}
T.~L. Marzetta, ``Noncooperative cellular wireless with unlimited numbers of base station antennas,'' {\em IEEE Transactions on Wireless Communications}, vol.~9, pp.~3590--3600, Nov. 2010.

\bibitem{Ngo2012}
H.~Q. Ngo and E.~G. Larsson, ``{EVD}-based channel estimation in multicell multiuser {MIMO} systems with very large antenna arrays,'' in {\em Proc. of IEEE International Conference on Acoustics, Speech and Signal Processing (ICASSP)}, 2012.

\bibitem{Yin2013}
H.~Yin, D.~Gesbert, M.~Filippou, and Y.~Liu, ``A coordinated approach to channel estimation in large-scale multiple-antenna systems,'' {\em IEEE Journal on Selected Areas in Communications}, vol.~31, no.~2, pp.~264--273, 2013.

\bibitem{Cottatellucci2013}
L.~Cottatellucci, R.~R. M{\"u}ller, and M.~Vehkapera, ``Analysis of pilot decontamination based on power control,'' in {\em Proc. of IEEE 77th Vehicular Technology Conference (VTC-Spring)}, 2013.

\bibitem{Mueller2014}
R.~R. M{\"u}ller, L.~Cottatellucci, and M.~Vehkapera, ``Blind pilot decontamination,'' {\em IEEE Journal of Selected Topics in Signal Processing}, vol.~8, no.~5, pp.~773--786, 2014.

\bibitem{Yin2016}
H.~Yin, L.~Cottatellucci, D.~Gesbert, R.~R. M{\"u}ller, and G.~He, ``Robust pilot decontamination based on joint angle and power domain discrimination,'' {\em IEEE Transactions on Signal Processing}, vol.~64, no.~11, pp.~2990--3003, 2016.

\bibitem{Bjoernson2020}
E.~Bj{\"o}rnson and L.~Sanguinetti, ``Making cell-free massive {MIMO} competitive with {MMSE} processing and centralized implementation,'' {\em IEEE Transactions on Wireless Communications}, vol.~19, no.~1, pp.~77--90, 2020.

\bibitem{Demir2021}
{\"O}.~T. Demir, E.~Bj{\"o}rnson, and L.~Sanguinetti, ``Foundations of user-centric cell-free massive mimo,'' {\em Foundations and Trends{\textregistered} in Signal Processing}, vol.~14, no.~3-4, pp.~162--472, 2021.

\bibitem{Wang2020}
H.~Wang, A.~Kosasih, C.-K. Wen, S.~Jin, and W.~Hardjawana, ``Expectation propagation detector for extra-large scale massive {MIMO},'' {\em IEEE Transactions on Wireless Communications}, vol.~19, no.~3, pp.~2036--2051, 2020.

\bibitem{Minka2001a}
T.~P. Minka, {\em A family of algorithms for approximate {B}ayesian inference}.
\newblock PhD thesis, Massachusetts Institute of Technology, 2001.

\bibitem{Minka2001b}
T.~P. Minka, ``Expectation propagation for approximate {B}ayesian inference,'' in {\em Proc. of 17th Conference on Uncertainty in Artificial Intelligence (UAI)}, pp.~362--369, 2001.

\bibitem{Cespedes2014}
J.~C{\'e}spedes, P.~M. Olmos, M.~S{\'a}nchez-Fern{\'a}ndez, and F.~Perez-Cruz, ``Expectation propagation detection for high-order high-dimensional {MIMO} systems,'' {\em IEEE Transactions on Communications}, vol.~62, no.~8, pp.~2840--2849, 2014.

\bibitem{Ghavami2017b}
K.~Ghavami and M.~Naraghi-Pour, ``{MIMO} detection with imperfect channel state information using expectation propagation,'' {\em IEEE Transactions on Vehicular Technology}, vol.~66, no.~9, pp.~8129--8138, 2017.

\bibitem{Ghavami2018}
K.~Ghavami and M.~Naraghi-Pour, ``Blind channel estimation and symbol detection for multi-cell massive {MIMO} systems by expectation propagation,'' {\em IEEE Transactions on Wireless Communications}, vol.~17, no.~2, pp.~943--954, 2018.

\bibitem{Ngo2020}
K.-H. Ngo, M.~Guillaud, A.~Decurninge, S.~Yang, and P.~Schniter, ``Multi-user detection based on expectation propagation for the non-coherent {SIMO} multiple access channel,'' {\em IEEE Transactions on Wireless Communications}, vol.~19, no.~9, pp.~6145--6161, 2020.

\bibitem{Zhang2020}
Z.~Zhang, H.~Li, Y.~Dong, X.~Wang, and X.~Dai, ``Decentralized signal detection via expectation propagation algorithm for uplink massive {MIMO} systems,'' {\em IEEE Transactions on Vehicular Technology}, vol.~69, no.~10, pp.~11233--11240, 2020.

\bibitem{Dong2022}
Y.~Dong, H.~Li, C.~Gong, X.~Wang, and X.~Dai, ``An enhanced fully decentralized detector for the uplink {M}-{MIMO} system,'' {\em IEEE Transactions on Vehicular Technology}, vol.~71, no.~12, pp.~13030--13042, 2022.

\bibitem{Li2023}
H.~Li, Y.~Dong, C.~Gong, X.~Wang, and X.~Dai, ``Decentralized groupwise expectation propagation detector for uplink massive {MU}-{MIMO} systems,'' {\em IEEE Internet of Things Journal}, vol.~10, no.~6, pp.~5393--5405, 2023.

\bibitem{Kosasih2021}
A.~Kosasih, V.~Miloslavskaya, W.~Hardjawana, V.~Andrean, and B.~Vucetic, ``Improving cell-free massive {MIMO} detection performance via expectation propagation,'' in {\em Proc. of IEEE 94th Vehicular Technology Conference (VTC-Fall)}, 2021.

\bibitem{He2021}
H.~He, H.~Wang, X.~Yu, J.~Zhang, S.~H. Song, and K.~B. Letaief, ``Distributed expectation propagation detection for cell-free massive {MIMO},'' in {\em Proc. of IEEE Global Communications Conference (GLOBECOM)}, 2021.

\bibitem{He2023}
H.~He, X.~Yu, J.~Zhang, S.~H. Song, and K.~B. Letaief, ``Cell-free massive {MIMO} detection: A distributed expectation propagation approach,'' {\em arXiv}, 2023.

\bibitem{Ghavami2017a}
K.~Ghavami and M.~Naraghi-Pour, ``Noncoherent {SIMO} detection by expectation propagation,'' in {\em Proc. of IEEE International Conference on Communications (ICC)}, 2017.

\bibitem{schulke2016statistical}
C.~Sch{\"u}lke, {\em Statistical physics of linear and bilinear inference problems}.
\newblock PhD thesis, Universit{\'e} Paris Diderot, Sapienza Universit{\`a} di Roma, 2016.

\bibitem{Minka2005}
T.~Minka, ``Divergence measures and message passing,'' Technical report MSR-TR-2005-173, 2005.

\bibitem{Heskes2005}
T.~Heskes, M.~Opper, W.~Wiegerinck, O.~Winther, and O.~Zoeter, ``Approximate inference techniques with expectation constraints,'' {\em Journal of Statistical Mechanics: Theory and Experiment}, vol.~2005, no.~11, p.~P11015, 2005.

\bibitem{Candan2021}
C.~Candan, ``Proper definition and handling of dirac delta functions [lecture notes],'' {\em IEEE Signal Processing Magazine}, vol.~38, no.~3, pp.~186--203, 2021.

\end{thebibliography}

\vfill\pagebreak

\end{document}